%% file: main_v1.tex
\documentclass[12pt]{article}

\usepackage{comment}
\usepackage[ruled,vlined]{algorithm2e}
\usepackage{algpseudocode}
\usepackage{amsmath}
\usepackage{amsfonts}
\usepackage{amssymb}
\usepackage{svg}
\usepackage{graphicx}
\usepackage{subcaption}
\usepackage{enumerate}
\usepackage{geometry}
\usepackage{tikz}
\usetikzlibrary{positioning,arrows.meta,calc}
\usepackage[colorlinks=true, allcolors=blue]{hyperref}
\usepackage[round]{natbib}
\setcitestyle{authoryear,round,aysep={}}
\usepackage{xcolor}
\usepackage{esvect}
\usepackage{caption}
\usepackage{setspace}
\usepackage{caption}
\usepackage{xcolor}
\usepackage[english]{babel}
\usepackage{amsthm}
\usepackage[toc,page]{appendix}
\usepackage{booktabs}
\usepackage{makecell}
\usepackage{booktabs}
\usepackage{threeparttable}
\usepackage[colorinlistoftodos,textsize=tiny,textwidth=50pt]{todonotes} 

\theoremstyle{plain}
\newtheorem{theorem}{Theorem}[section]

\newtheorem{corollary}[theorem]{Corollary}

\newtheorem{assumption}{Assumption}
\newtheorem{example}{Example}

\theoremstyle{definition}

\theoremstyle{remark}

\title{Structural Fine-Tuning under Economic Restrictions\thanks{We would like to thank Karun Adusumilli, Xu Cheng, Francis Diebold, Wayne Gao, Joao Gomes, Sasha Indarte, Benjamin Keys, Dirk Krueger, Victor Orestes, Nikolai Roussanov, Frank Schorfheide, Xiaoxia Shi, Petra Todd, and Zhijie Xiao, as well as participants at the University of Pennsylvania Econometrics Lunch, for their helpful comments, guidance, and feedback on this project. We gratefully acknowledge internal Wharton Research Data Services (iWRDS) for providing data and infrastructure support.}}
\author{Yi Niu\thanks{Department of Economics, University of Pennsylvania. Email: \href{mailto:n1y1@sas.upenn.edu}{\texttt{n1y1@sas.upenn.edu}}.}\and Yuecheng Su\thanks{\raggedright Finance Department, The Wharton School, University of Pennsylvania. Email: \href{mailto:suyc@wharton.upenn.edu}{\texttt{suyc@wharton.upenn.edu}}.}\and Ethan Yang\thanks{Department of Economics, University of Pennsylvania. Email: \href{mailto:etyang17@sas.upenn.edu}{\texttt{etyang17@sas.upenn.edu}}.}}
\date{}

\begin{document}
\maketitle

\begin{abstract}
How should a predictor trained on historical data be adapted to a new economic environment or task when few target observations are available? We propose structural fine-tuning, a framework that introduces economic restrictions to guide the adaptation of pretrained predictors. The framework does not require a complete structural model; instead, it utilizes constraints based on economic knowledge, ranging from simple, ubiquitous assumptions to particular workhorse models. Theoretically, we characterize conditions when economic restrictions improve prediction over the original model and ordinary fine-tuning. Then, we illustrate the framework through a simulation exercise of a credit issuer's credit-limit decision following a Medicaid-style expansion in medical payments. Imposing restrictions based on eligibility, outstanding medical bills, cash flows, and group-level risk bounds, our framework lowers the root mean squared error of predicted risk by 28 to 39 percent and the issuer's policy regret by 3 to 10 percent relative to ordinary neuron gating, LoRA, and full fine-tuning.
\end{abstract}

\newpage
\doublespacing

\section{Introduction}
How should a predictor trained on historical data be adapted to a new economic environment or task when few target observations are available?  Policy reforms, changes in market institutions, or aggregate shocks can change both the composition of a population and the relationship between its characteristics and outcomes. A research team or institute, such as The Federal Reserve, may have a source predictor from a large historical data but only a small sample from the new target environment. While the source predictor likely carries useful information, adapting it a new environment is challenging with little target data. Therefore, researchers have developed various fine-tuning methods to efficiently update the source model. For economists, another common approach is to structurally model the target environment using economic knowledge. 

In this paper, we combine the strengths of those two approaches by studying how a collection of economic restrictions, ranging from simple, commonsensical rules to specific models, improves fine-tuning. We ask three questions. When does adaptation improve on leaving the source predictor unchanged? When does introducing economic restrictions improve on adaptation without them? How do these comparisons depend on the accuracy of the maintained economic restrictions?

These questions matter because large pretrained models are increasingly used in economic measurement and forecasting \citep{dell2025deep,ludwig2026large}. However, a prediction model that fits historical data well in the source environment may not do so in the target environment. Reproducing those large-scale pretraining models using target data can be prohibitively costly. Even when retraining is feasible, a large historical data does not provide the same information as the few target data. Therefore, adapting an existing model offers a practical alternative \cite{hu2022lora}. Nonetheless, traditional fine-tuning methods leaves out valuable economic knowledge. Our framework aims to fill this gap by introducing economic restrictions.

The selection of economic restrictions also matters. In principal, our framework can accommodate restrictions from almost any economic models. However, for a given economic question, economists may hold different beliefs on structural conditions, and there may not exist a universally accepted model. Still, they can agree on particular assumption, such as bounds on the response, monotonicity, or concavity in the inputs. These agreements, which we call ubiquitous economic restrictions (UER), can help fine-tuning by distinguishing among predictors that target observations cannot reliably separate while preserving the flexibility of nonparametric estimation.

Existing work offers two complementary ways to guide adaptation. Transfer-learning methods control which parameters change and how far the fitted model moves from its source initialization \citep{swietojanski2014learning,li2018explicit,hu2022lora}. Shape-restricted estimations use substantive information about the admissible response \citep{matzkin1994restrictions,chetverikov2017nonparametric}. However, limiting parameter changes does not ensure that the updated response utilizes economic restrictions, and satisfying those restrictions does not ensure that updating improves prediction accuracy. The open question is what restrictions on the target response add. To assess the value of these economic restrictions, we need to separate the value of updating the source predictor from the gains and costs of introducing these restrictions.

In this paper, we develop a framework called \emph{structural fine-tuning} that unifies the fine-tuning with the economic restrictions. Economists choose a collection of economic models related to the target question and identify their UERs. The adaptation method defines a family of predictors that includes the source predictor. Structural fine-tuning uses the target sample to fit a predictor within the subset that satisfies the UER, while penalizing departure from the source predictions. Without these restrictions, it becomes an ordinary fine-tuning using the same loss and penalty. Hence, we use the target data to choose among updates that satisfy the shared restrictions, without committing to a single economic model. This framework works with a broad class of fine-tuning methods, including neuron gates, Low-Rank Adaptation (LoRA), and full fine-tuning.

We derive theoretical results for last-layer neuron gating, which makes the fine-tuned predictor an affine function of the gates. For economic restrictions, we derive the expected gains from structural fine-tuning relative to both source predictor and ordinary fine-tuning. We decompose the expected gain into contributions from changes in squared bias and variance. For example, misspecified restrictions can add bias, and estimated restrictions may increase variance. A bound on the deviation of the true response from the economic restrictions yields a sufficient condition for structural fine-tuning to improve on both alternatives in expectation. Under valid convexity and regularity conditions discussed in Section \ref{sec:theory-convex-restrictions}, structural fine-tuning yields weakly lower prediction error than ordinary fine-tuning. 

Our credit market simulation draws on the Medicaid expansion under the \textit{Affordable Care Act}, which increase the credit card borrowing mainly by expanding credit supply \citep{bornstein2026impact}. To decide whose credit limit to raise, a lender updates its model using data from 1,000 post-expansion borrowers. The lender imposes seven economic restrictions. Risk stays at its historical level if a borrower is ineligible or has no qualifying medical bill. For borrowers who receive a payment, only historical risk and relevant cash flows can capture the effect of borrower history with a small tolerance. The other UERs bound the delinquency risks at group-level. We run this with neuron gates, LoRA, and full fine-tuning, both with and without the seven restrictions. Imposing the restrictions cuts the root mean squared error (RMSE) of predicted risk by approximately 28 to 39 percent. The reduction in average policy regret relative to ordinary fine-tuning is from 3.13 to 10.21 percent across the three methods. In each simulation draw, the lender also earns more with each restricted fit than it would by continuing the historical model.

More broadly, our findings suggest that economic theory and machine learning play complementary roles when the economy changes. Pretrained models can learn complex relationships in historical data, but the historical sample alone cannot establish which relationships still hold. Economic theories can guide this update by identifying restrictions that are valid in the new environment. For example, considering policy changes, this provides a balanced response to the concern that relationships estimated under one policy might break down under another \citep{lucas1976econometric}. As researchers keep adapting existing models, they need to decide how much theory to impose on each update. Our framework can make this decision more explicit by assessing the gains and costs of each economic restriction. While it does not replace structural model's role in counterfactual analysis, structural fine-tuning offers theory a role in adapting predictors.

\paragraph{Related Literature} This paper contributes to the literature on statistical transfer learning and forecasting under instability by incorporating insights from economically-motivated machine learning and estimation with economic restrictions. The concern that historical relationships may break down after a policy change is central to \cite{lucas1976econometric}. \cite{stock1996evidence} documents widespread instability in macroeconomic forecasting that adaptive methods struggle to capture. The most common problem is how to retain the precision from older information without additional bias. Forecasting methods address it by changing the estimation window and the weights placed on observations from different regimes \citep{pesaran2007selection, pesaran2013optimal}. On the other hand, high-dimensional transfer learning addresses a related problem by restricting how much source and target coefficients can differ\citep{bastani2021predicting, li2022transfer}. An informative shared representation can also reduce the complexity of learning a new task \citep{tripuraneni2020theory}. What we gain in these settings depend on the relationship between environments we assumed. We complement this work by imposing economic restrictions on some target responses within the selected adaptation family. We study task-specific complexity in terms of those restrictions, without invoking general representation-learning guarantees. The target data must still be informative enough to estimate the response. What the restrictions can provide is a narrower set of adjustments, allowing the model to avoid the full update when there is an economic shift. 

The second branch of literature treats economic theory as information for estimation. \cite{toro1968test} found that the trade-off between lower variance and bias comes from incorrect restrictions. Shape restriction applies this idea to nonparametric responses \citep{matzkin1994restrictions,blundell2012measuring,chetverikov2017nonparametric,chetverikov2018econometrics}. Here, we don't need to impose theory as an exact restriction. Priors derived from structural models and shrinkage theory-consistent parameter sets let the data determine the strength of regularization \citep{delnegro2004priors,fessler2019how}. Restrictions can also be applied after estimation: rearrangement and shape-restricted operators work with generic preliminary estimates, including predictors obtained through transfer learning \citep{chernozhukov2009improving,chen2021shape}. Therefore, what distinguishes us from the previous literature is not which stage economic restrictions are imposed. We fit within the economically admissible response set and a specified adaptation family. We use the target data to select an update that satisfies the economic restrictions and stays within the specified adaptation family. A generic shape transformation of a fitted predictor does not need to stay within that family. We extend the classical restricted-estimation tradeoff to the adaptation of pretrained predictors under distribution shift. The question is how much information to keep from the source model and how much to learn from the limited target data. Using an exactly solvable finite-support response-space benchmark, we evaluate economically constrained neuron-gate adaptation and conventional fine-tuning against freezing and equally informed calibration methods.

The third literature imposes the economic structure into machine learning. Some approaches build monotonicity or convexity directly into the network architecture \citep{dugas2009incorporating, you2017deep}. For example, they can enter through estimation objectives such as no-arbitrage conditions in deep-learning asset pricing \citep{chen2024deep} or through pretraining data simulated from a structural model \citep{chen2026teaching}. In each case, economic theories enter during training. Fine-tuning methods constrain the following update via neuron gates, low-rank corrections, or regularization toward pretrained parameters \citep{swietojanski2014learning,hu2022lora,li2018explicit}. Here, the choice of the updated family matters a lot because previous literature \citep{kumar2022finetuning,lee2023surgical} shows that the full fine-tuning can distort representations out of distribution. However, the parameter efficiencies cannot guarantee the economic feasibility. We use the selected economic information to restrict the target response during the adaptation, which doesn't require either a structurally generated source sample or a single fully specified model. For instance, under the no-arbitrage condition, the European call prices are nonincreasing and convex in strike, but these restrictions do not specify a pricing formula \citep{merton1973theory}. We use the risk comparisons to show when economic restrictions can help an update. These results do not imply that our method dominates theory-guided training, and the two methods can be used together.

The rest of the paper proceeds as follows. Section~\ref{sec:methodology} introduces structural fine-tuning. Section~\ref{sec:theory} characterizes when it improves the prediction under last-layer neuron gating. Section~\ref{sec:simulation} presents the credit market simulation, and Section~\ref{sec:conclusion} concludes.

\section{Methodology}\label{sec:methodology}

The gist of fine-tuning is: researchers pretrained an estimator on a large source dataset, and they believe the estimator can capture relationships that are also useful for the target task. Then, fine-tuning starts from this pretrained estimator and uses the relatively smaller target observations to adapt it. In economics, researchers would have their believed structural models for the target data with a given question. We derive \textit{economic restrictions} (ER) from these beliefs and models. Imposing ER can rule out predictors that fit the sample well enough to be selected by ordinary fine-tuning but violate implications shared by the economic models, while maintaining the flexibility of nonparametric estimation. Our framework thus guides fine-tuning methods using economic restrictions derived from beliefs of the economic structure, hence the term ``structural fine-tuning." 

Nonetheless, these beliefs can be strong or weak over partially or fully specified structural models, but there will mostly be a nonempty intersection of these beliefs for this particular question. We call the structural assumptions lies in this intersection \emph{ubiquitous} economic restrictions (UER). For example, diminishing marginal utility can be considered as ubiquitous when studying household spending on food.  For the rest of the paper, we will focus on UER and  to avoid misspecification and illustrate universality of our results.

Section~\ref{sec:methodology-fine-tuning}  overviews two common fine-tuning methods. Then, Section~\ref{sec:methodology-structural-fine-tuning} formally introduces structural fine-tuning and explain how it connects to conventional fine-tuning methods.

\subsection{Fine-tuning} \label{sec:methodology-fine-tuning}

Let \(Y\in\mathbb R\) be the outcome and \(X\in\mathcal X\) a vector of \(k\) covariates, where \(\mathcal X\subseteq\mathbb R^k\) is measurable. Write \(P_S\) and \(P_T\) for the source and target laws of \((X,Y)\), which may coincide. Given a predictor \(\hat f_S\) trained on a large source dataset, we use \(n\) target observations \(\mathcal D_T=\bigl((X_j,Y_j)\bigr)_{j=1}^n\) to adapt it to the target function \(m_T:\mathcal X\to\mathbb R\). Denote \(\mathcal F\) as the class of measurable candidate target functions \(f:\mathcal X\to\mathbb R\). We adapt \(\hat f_S\) within a family of predictors \(\mathcal F_U=\{f_\theta:\theta\in\Theta\}\subseteq\mathcal F\). Here, \(\theta\) contains the parameters fitted on the target sample, and \(\Theta\) is their parameter space. The researcher-specified fine-tuning method specifies how \(\theta\) changes the source predictor. We assume that \(f_{\theta_S}=\hat f_S\) for some \(\theta_S\in\Theta\).

To fix ideas, let \(\hat f_S\) be a feedforward neural network with \(L\) hidden layers and \(d_\ell\) neurons in hidden layer \(\ell\). The input and output dimensions are \(d_0=k\) and \(d_{L+1}=1\). Define \(W_\ell^S\in\mathbb R^{d_\ell\times d_{\ell-1}}\) and \(b_\ell^S\in\mathbb R^{d_\ell}\) as the weights and biases fitted on the source data for \(\ell=1,\ldots,L+1\). Let \(\sigma_\ell\) denote the componentwise activation function in hidden layer \(\ell\) and \(\rho:\mathbb R\to\mathbb R\) the output link. With \(h_0(x)=x\), we write the source neural network as
\[
 \begin{aligned}
  h_\ell(x)&=\sigma_\ell\bigl(W_\ell^S h_{\ell-1}(x)+b_\ell^S\bigr),
  &&\ell=1,\ldots,L,\\
  \hat f_S(x)&=\rho\bigl(W_{L+1}^S h_L(x)+b_{L+1}^S\bigr).
 \end{aligned}
\]

\noindent We now introduce two commonly used fine-tuning methods.

\emph{Neuron gates.} Neuron gating adapts a neural network by learning a scalar multiplier for each selected neuron's activation \citep{swietojanski2014learning}. Define \(g_\ell\in[0,\infty)^{d_\ell}\) as the gate vector for hidden layer \(\ell\). Let \(\mathcal L_G\subseteq\{1,\ldots,L\}\) denote the selected hidden layers. Define
\[
 \theta=(g_\ell)_{\ell\in\mathcal L_G},
 \qquad \Theta=\prod_{\ell\in\mathcal L_G}[0,\infty)^{d_\ell}.
\]
 We set gates $ = 1$ in unselected hidden layers and hold \(W_\ell^S\) and \(b_\ell^S\) fixed. Let \(h_\ell^g(x)\) denote the gated neurons and \(\odot\) componentwise multiplication. With \(h_0^g(x)=x\), we write the gated neural network as
\[
 \begin{aligned}
  h_\ell^g(x)&=g_\ell\odot\sigma_\ell\bigl(W_\ell^S h_{\ell-1}^g(x)+b_\ell^S\bigr),
  &&\ell=1,\ldots,L,\\
  f_\theta(x)&=\rho\bigl(W_{L+1}^S h_L^g(x)+b_{L+1}^S\bigr).
 \end{aligned}
\]
A gate $g_\ell=0$ turns down that neuron's contribution to the next layer, $0 < g_\ell <1$ attenuates it, and $g_\ell > 1$ amplifies it. Setting every gate to one recovers \(\hat f_S\). The advantage of gating methods is computational. It fits \(\sum_{\ell\in\mathcal L_G}d_\ell\) parameters, so a large network can be adapted with far fewer trainable parameters than full fine-tuning. Moreover, if one restricts $g_\ell \in [0,1]$ and $\sum g_\ell = 1$, gates have a nice convex combination interpretation for the neurons they attached to.

\emph{Low-rank adaptation (LoRA).} LoRA adapts a network by fitting low-rank corrections to selected source weight matrices \citep{hu2022lora}. Let \(\mathcal L_{\mathrm{LoRA}}\subseteq\{1,\ldots,L+1\}\) denote the selected layers. For each \(\ell\in\mathcal L_{\mathrm{LoRA}}\), let \(1\leq r_\ell\leq\min\{d_{\ell-1},d_\ell\}\) be the pre-specified maximum rank for the correction to \(W_\ell^S\) . Define factor matrices \(A_\ell\in\mathbb R^{r_\ell\times d_{\ell-1}}\) and \(B_\ell\in\mathbb R^{d_\ell\times r_\ell}\), and 
\[
 \theta=\bigl((A_\ell,B_\ell)\bigr)_{\ell\in\mathcal L_{\mathrm{LoRA}}},
 \qquad
 \Theta=\prod_{\ell\in\mathcal L_{\mathrm{LoRA}}}
 \left(\mathbb R^{r_\ell\times d_{\ell-1}}
 \times\mathbb R^{d_\ell\times r_\ell}\right).
\]
 We estimate \(A_\ell\) and \(B_\ell\) in the selected layers while holding \(W_\ell^S\) and \(b_\ell^S\) fixed for all $\ell$. Define \(W_\ell^\theta\) as the adapted weight matrix. In each selected layer, write
\[
 W_\ell^\theta=W_\ell^S+B_\ell A_\ell,
 \qquad \operatorname{rank}(B_\ell A_\ell)\leq r_\ell.
\]
In the unselected layers, \(W_\ell^\theta=W_\ell^S\). Intuitively, the columns of \(B_\ell\) specify directions for the correction, and \(A_\ell\) determines how the layer's inputs are combined with these directions. The advantage of LoRA comes from its low-rank property. We only need \(r_\ell(d_{\ell-1}+d_\ell)\) parameters in layer \(\ell\), which can be far fewer than the \(d_\ell d_{\ell-1}\) entries in its weight matrix when the rank $r_\ell$ is small. Also, since \(B_\ell A_\ell\) can be merged into \(W_\ell^S\), LoRA preserves the original network architecture.

\begin{example}[Conditional mean with last-layer gates]\label{ex:method-last-layer-gates}
Suppose the target estimand is \(m_T(x)=\mathbb E_{P_T}[Y\mid X=x]\), and researchers estimated $\mathbb E_{P_S}[Y\mid X=x]$ by a NN with a linear output \(\rho(t)=t\) on the source data. Researchers may decide to use last-layer gates to do fine-tuning, which attaches gates only to the last hidden layer, so \(\mathcal L_G=\{L\}\). Compactly write \(g=g_L\in[0,\infty)^{d_L}\), \(a^S=(b_{L+1}^S)_1\) as the source output intercept, and \(w^S=(W_{L+1}^S)^\top\in\mathbb R^{d_L}\) as the source output weights. Holding all source weights and biases fixed, then we can write the fine-tuned predictor as
\[
 f_g(x)=a^S+(w^S)^\top\bigl(g\odot h_L(x)\bigr).
\]
Notice that \(f_g\) is affine in \(g\)\footnote{Refitting a linear output layer while keeping hidden features fixed is commonly called \emph{linear probing} (\href{https://arxiv.org/abs/2202.ten54}{Kumar et al., 2022}). Nonnegative gates give a restricted linear output layer, with coefficients \(w_j^Sg_j\) and a fixed intercept. Throughout the paper, we use \emph{fine-tuning} broadly to include linear probing.} . We can therefore estimate the gates by solving a nonnegative least-squares problem
\[
 \hat g\in
 \underset{g\in[0,\infty)^{d_L}}{\arg\min}
 \frac{1}{n}\sum_{j=1}^n
 \bigl\{Y_j-a^S-(w^S)^\top\bigl(g\odot h_L(X_j)\bigr)\bigr\}^2.
\]
For other gating architectures, the predictor can depend nonlinearly on the gates, and this reduction to nonnegative least squares need not hold.
\end{example}

\subsection{Structural Fine-tuning} \label{sec:methodology-structural-fine-tuning}

\begin{figure}[!htb]\centering\includegraphics[width=0.8\textwidth]{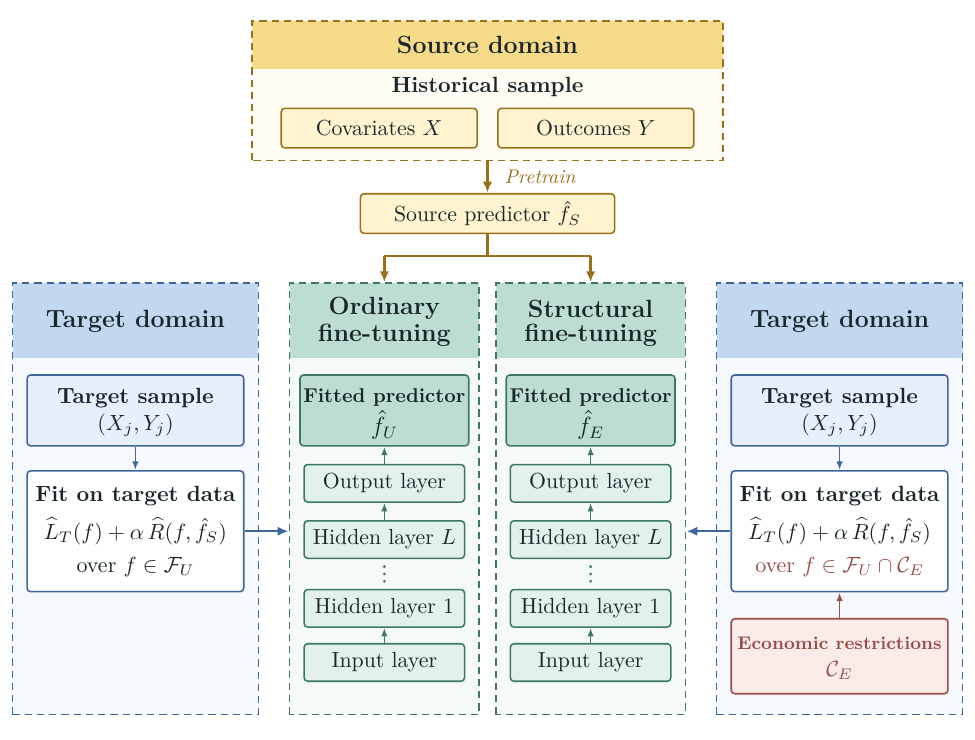}\caption{Ordinary fine-tuning (left) and structural fine-tuning (right).}\label{fig:fine-tuning-comparison}\end{figure}

Economic intuition can guide fine-tuning by specifying restrictions that the object of interest should satisfy. For example, an economic model may imply that the function is monotone, concave, or convex in a given input. These shape restrictions can rule out functional forms when fitting the target data. We call fine-tuning \emph{structural} when we impose these economic restrictions as constraints in the fine-tuning optimization problem.

There are three common motivations to impose ER during fine-tuning. First, the source model may have been trained for a broad purpose. Therefore, the researcher may not have incorporated ER for a particular downstream task during the source training stage. When the researcher later returns to this task with new data, they can impose the corresponding ER during fine-tuning. Second, the source and target distributions may differ while some ER remain valid and can guide adaptation to the target population. Third, with limited target data, several functions may fit the observed outcomes similarly well; ER helps select functions for the fine-tuning problem. In each case, we focus on UER shared by the economic models under consideration. We impose these common implications while using the target data to fine tune the target function of interest. We avoid stronger restrictions because they can undermine the flexibility that motivates nonparametric estimation and make the estimate more sensitive to model misspecification.

Now we formally introduce structural fine-tuning. Let \(\mathcal C_E\) be the set of candidate functions that satisfy the maintained UER.\footnote{Some applications involve additional parameters or functions in the loss or UER. Known quantities and existing estimates, such as a bound estimated from source data, are held fixed during fine-tuning. Unknown nuisance quantities may instead be estimated jointly with \(f\). } For a candidate function \(f\), let \(M_E(f)\) and \(I_E(f)\) denote the equality and inequality constraints. The UER constraints define the set of admissible functions
\begin{equation}\label{eq:method-economic-set}
 \mathcal C_E
 =\{f\in\mathcal F:M_E(f)=0,\quad I_E(f)\leq 0\}.
\end{equation}
We assume \(m_T\in\mathcal C_E\).

We substitute \(f_\theta\) for \(f\) during structural fine-tuning. The feasible parameter set is
\begin{equation}\label{eq:method-feasible-parameters}
 \Theta_E
 =\{\theta\in\Theta:M_E(f_\theta)=0,\quad I_E(f_\theta)\leq0\}.
\end{equation}
The fine-tuning method determines the map \(\theta\mapsto f_\theta\), and the economic model determines the restrictions on its predictions. Therefore, we can impose ER upon neuron gates, LoRA, full fine-tuning, and many other fine-tuning methods.

Let \(\widehat L_T(f)\) be the empirical loss for a candidate function \(f\). For a target function defined by moment conditions, the loss can measure the discrepancy between empirical moments and the moments implied by \(f\). Let \(\widehat R(f,\hat f_S)\) be a nonnegative penalty for departing from the source predictor, with \(\widehat R(\hat f_S,\hat f_S)=0\).

For a given penalty weight \(\alpha\geq0\), the structural fine-tuning optimization problem is
\begin{equation}\label{eq:method-estimator}
 \hat f_E\in
 \underset{f\in\mathcal F_U\cap\mathcal C_E}{\arg\min}
 \left\{\widehat L_T(f)+\alpha\widehat R(f,\hat f_S)\right\}.
\end{equation}
Here \(\hat f_E\) is the estimated target function. The intuition for the penalty term is that the source predictor is trained on a large sample, and we expect it to be informative for the target task. We include \(\alpha\widehat R(f,\hat f_S)\) to discourage large departures from the source predictor during fine-tuning. The penalty \(\widehat R\) specifies how departure is measured, while \(\alpha\) determines the strength of the penalization.

We illustrate \eqref{eq:method-estimator} with a concrete example.

\begin{example}[Household food shares]\label{ex:method-household}
Let \(X=(B,W)\) collect budget \(B>0\) and household history \(W\). Suppose food share is \(Y=\min\{1,V(1+N/B)\}\), with latent taste \(V\in(0,1)\) and need \(N\geq0\). Among households with the same history, the mix of tastes and needs is stable across budgets. We want to estimate \(m_T(b,w)=\mathbb E_{P_T}[Y\mid B=b,W=w]\). Write \(X_j=(B_j,W_j)\) and \(b_\lambda=\lambda b+(1-\lambda)b'\). With squared loss and a squared source penalty, we solve
\begin{equation}\label{eq:method-household-problem}
\begin{aligned}
 \hat f_E\in\underset{f\in\mathcal F_U}{\arg\min}\quad
 &\frac{1}{n}\sum_{j=1}^n
 \bigl[\{Y_j-f(X_j)\}^2+\alpha\{f(X_j)-\hat f_S(X_j)\}^2\bigr]\\
 \text{subject to}\quad
 &0\leq f(b,w)\leq1,\qquad f(b',w)\leq f(b,w),\\
 &b f(b,w)\leq b'f(b',w),\\
 &b_\lambda f(b_\lambda,w)
 \geq\lambda b f(b,w)+(1-\lambda)b'f(b',w).
\end{aligned}
\end{equation}
Here, \(0\leq\lambda\leq1\) and \(b'>b>0\), with the constraints imposed over the budgets and histories of interest.
\end{example}
In Example \ref{ex:method-household}, the ER are requiring predicted food shares non-increasing and predicted food spending non-decreasing and concave in budget. We believe these are ubiquitous ER across most household studies. 

Importantly, it is researchers' freedom to choose how ubiquitous their set of ER is. It is reasonable to impose more structural and less ubiquitous ER based on researchers' believed class of models. Back to Example \ref{ex:method-household}, a real empirical research would reasonably argue that the food-share curves for some demographic groups can be modeled with a common shape. \citet{blundell2007semi} estimate a common Engel curve while allowing the expenditure scale and the level of the budget share to differ across groups. To impose this semi-ubiquitous ER (SUER), we require that there exist a budget multiplier \(s>0\) and a food-share shift \(c\in\mathbb R\) such that $ f(sb,w_1)=f(b,w_0)+c$ for every \(b\) in a given range for two household histories \(w_0\) and \(w_1\). Therefore, how ubiquitous researchers want their ER to be depends on how much they trust their class of models, given the risk of misspecification.

We implement \eqref{eq:method-estimator} using the following algorithm. Write \(\widehat L_{\rm val}\) for the target loss evaluated on the validation sample.

\begingroup
\begin{singlespace}
\begin{algorithm}[H]
\small
\DontPrintSemicolon
\LinesNumbered
\caption{Structural fine-tuning}\label{alg:structural-fine-tuning}
\KwIn{Source fit \(\hat f_S\), fine-tuning family \(\mathcal F_U\), target data \(\mathcal D_T\), UER \(\mathcal C_E\), and a nonempty finite grid \(\mathcal A\subset[0,\infty)\) of penalty weights.}
\KwOut{Fitted function \(\hat f_E\) and selected weight \(\hat\alpha\), or \emph{no accepted fit}.}
Split \(\mathcal D_T\) into nonempty training and validation samples \(\mathcal D_{\rm tr}\) and \(\mathcal D_{\rm val}\)\;
\For{\(\alpha\in\mathcal A\)}{
  Use \(\mathcal D_{\rm tr}\) to fit \(\hat f_\alpha\) under \eqref{eq:method-estimator}\;
}
Let \(\mathcal A_{\rm ok}\) contain the weights for which a fit \(\hat f_\alpha\in\mathcal C_E\) is obtained\;
\If{\(\mathcal A_{\rm ok}=\varnothing\)}{
  \Return{\emph{no accepted fit}}\;
}
Select \(\displaystyle\hat\alpha\in\arg\min_{\alpha\in\mathcal A_{\rm ok}}\widehat L_{\rm val}(\hat f_\alpha)\)\;
Refit \eqref{eq:method-estimator} on all of \(\mathcal D_T\) at \(\hat\alpha\) to obtain \(\hat f_E\)\;
\If{no refit \(\hat f_E\in\mathcal C_E\) is obtained}{
  \Return{\emph{no accepted fit}}\;
}
\Return{\((\hat f_E,\hat\alpha)\)}\;
\end{algorithm}
\end{singlespace}
\endgroup

\section{Theory}\label{sec:theory}

When does structural fine-tuning improve prediction relative to frozen source model or ordinary fine-tuning? We study these comparisons using the affine representation of last-layer gating with a linear output, as in Example~\ref{ex:method-last-layer-gates}. The target function is the conditional mean, \(m_T(x)=\mathbb E_{P_T}[Y\mid X=x]\).

Section~\ref{sec:theory-setup} formally establishes our theoretical framework. Section~\ref{sec:theory-prediction-improvement} characterizes sufficient conditions under which structural fine-tuning improves on ordinary fine-tuning and frozen source model. Finally, Section~\ref{sec:theory-convex-restrictions} characterizes bounds on the gains over frozen source model.

\subsection{Setup}\label{sec:theory-setup}

Let \(Q\) be a fixed, known evaluation distribution with \(m_T\) specified \(Q\)-almost everywhere. With \(\|v\|_Q^2=\int v^2\,dQ\), define
\begin{equation}\label{eq:theory-prediction-error}
 \mathcal R(f)=\|f-m_T\|_Q^2.
\end{equation}
When \(Q=P_T^X\), the target marginal law of \(X\), and \(\mathbb E_{P_T}[Y^2]<\infty\), this is excess mean squared prediction error relative to \(m_T\). We compare the frozen fit \(\hat f_S\) with ordinary and structural fits \(\hat f_U,\hat f_E\), which minimize \eqref{eq:method-estimator} over \(\mathcal F_U\) and \(\mathcal F_E=\mathcal F_U\cap\mathcal C_E\) at a common \(\alpha\).

Using the notation of Example~\ref{ex:method-last-layer-gates}, let \(p=d_L\), let \(h_{Lj}(x)\) be the \(j\)-th entry of \(h_L(x)\), and define \(\phi_j(x)=w_j^S h_{Lj}(x)\). With \(\beta=g-\mathbf1_p\), where \(\mathbf1_p\) is the vector of ones, last-layer gating has the affine form
\[
 f_g(x)=\hat f_S(x)+\phi(x)^\top\beta=:f_\beta(x).
\]
This keeps the source representation fixed, as in linear probing \citep{tian2020rethinking,du2021fewshot,kumar2022finetuning}. For Theorem~\ref{thm:theory-expected-gains}, we allow \(\beta\in\mathbb R^p\), which permits signed gates. With the nonnegative gates in Example~\ref{ex:method-last-layer-gates}, \(\beta\geq-\mathbf1_p\). We impose the same bounds on both fits in Corollary~\ref{cor:theory-convex-restrictions}.

Let \(\beta_*\) minimize \(\mathcal R(f_\beta)\) over \(\mathbb R^p\), choosing the minimum-Euclidean-norm coefficient in case of a tie. Let \(h=m_T-f_{\beta_*}\) be the approximation error, \(M=\int\phi\phi^\top\,dQ\) the second-moment matrix of the features, and \(\|a\|_M^2=a^\top Ma\). Orthogonality \(\int\phi h\,dQ=0\) gives

\begin{equation}\label{eq:theory-affine-risk}
 \mathcal R(f_\beta)=\|h\|_Q^2+\|\beta-\beta_*\|_M^2.
\end{equation}

Let \(T\) contain \(q\) linear functionals, where \(1\leq q\leq p\). The economic restrictions require their values to equal \(\ell\in\mathbb R^q\). Their common domain is a linear space containing \(\hat f_S,m_T,\phi_1,\ldots,\phi_p\). With \(M_E(f)=T(f)-\ell\), the economic equalities for \(f_\beta\) are
\[
 A\beta=b,\qquad A_{ij}=T_i(\phi_j),\qquad b=\ell-T(\hat f_S).
\]
For the best affine approximation, let
\begin{equation}\label{eq:theory-economic-gap}
 d=b-A\beta_*=\ell-T(m_T)+T(h).
\end{equation}
An error in the required economic values contributes \(\ell-T(m_T)\) to \(d\); approximating \(m_T\) adds \(T(h)\). 

Let \(D\in\mathbb R^{n\times p}\) have entries \(D_{ij}=\phi_j(X_i)\). Subtracting the source prediction from each target outcome gives \(Z\in\mathbb R^n\), with \(Z_i=Y_i-\hat f_S(X_i)\). Let \(\alpha\geq0\) and \(K\in\mathbb R^{p\times p}\) be the penalty weight and matrix. The ordinary fit minimizes
\begin{equation}\label{eq:theory-coefficient-criterion}
 \frac1n\|Z-D\beta\|_2^2+\alpha\beta^\top K\beta
\end{equation}
over \(\mathbb R^p\), and the structural fit minimizes it over \(\{\beta:A\beta=b\}\). Let \(\hat\beta_U,\hat\beta_E\) be the exact minimizers, with fitted functions \(\hat f_j=f_{\hat\beta_j}\), \(j\in\{U,E\}\).

Suppose we want to keep the fine-tuned predictions close to the source predictions at the observed target inputs. We can penalize the average squared change
\[
 \frac1n\sum_{i=1}^n\{f_\beta(X_i)-\hat f_S(X_i)\}^2
 =\beta^\top G\beta,\qquad G=D^\top D/n.
\]
This gives \(K=G\). If we average the squared change under \(Q\), we get \(\|f_\beta-\hat f_S\|_Q^2=\beta^\top M\beta\) and use \(K=M\). For any \(K\), the quadratic term in \eqref{eq:theory-coefficient-criterion} is \(\beta^\top H\beta\), where \(H=G+\alpha K\).

Let \(\mathcal G=\sigma(\hat f_S,\phi,X_1,\ldots,X_n,A,b,K,\alpha)\). We compare the risks conditional on this information. Define \(h_X=(h(X_i))_{i=1}^n\) as the approximation errors at the fitting inputs and \(\xi=(Y_i-m_T(X_i))_{i=1}^n\) as the target errors. Thus \(Z=D\beta_*+h_X+\xi\).

\subsection{When does structural fine-tuning improve prediction?} \label{sec:theory-prediction-improvement}

All assumptions below are for fixed \(Q\) and hold almost surely under the joint sampling law.

\begin{assumption}[Regularity and sampling]\label{ass:theory-regularity}
Suppose \(\hat f_S,m_T,\phi_1,\ldots,\phi_p\in L^2(Q)\), \(A\) has rank \(q\), \(K\) is symmetric positive semidefinite, and \(H\) is positive definite. The target errors have conditional moments
\begin{equation}\label{eq:theory-design-noise}
 \mathbb E[\xi\mid\mathcal G]=0,\qquad
 \operatorname{Var}(\xi\mid\mathcal G)=\Sigma,
\end{equation}
where all entries of \(\Sigma\) are finite.
\end{assumption}

Full row rank of \(A\) guarantees that we can solve \(A\beta=b\). Positive definite \(H\) then gives unique fits, whose prediction errors are finite by square-integrability. We can take \(K=I_p\) and \(\alpha>0\) to get \(H\succ0\), even with \(p>n\).

\begin{assumption}[Economic equalities]\label{ass:theory-equalities}
We impose \(T(f_\beta)=\ell\) on the affine family, or equivalently \(A\beta=b\), with \(A\) and \(b\) defined in Section~\ref{sec:theory-setup}.
\end{assumption}

For example, suppose an independent expenditure audit reports a group's total spending. We may require predicted spending to match this total. If the source fit gives a different total, \(b\ne0\). The discrepancy \(d\) includes error in the audited total and approximation error for \(m_T\).

Define the conditional bias and covariance of the ordinary coefficient estimate by
\[
 e=H^{-1}\{D^\top h_X/n-\alpha K\beta_*\},\qquad
 V_U=H^{-1}D^\top\Sigma D H^{-1}/n^2.
\]
The equalities enter through
\begin{align*}
 J&=H^{-1}A^\top(AH^{-1}A^\top)^{-1},&
 P&=I_p-JA.
\end{align*}
Here \(I_p\) is the \(p\)-dimensional identity matrix. For \(v\in\mathbb R^q\), \(Jv\) solves \(Aa=v\) with the least \(a^\top Ha\), while \(P\) projects coefficient updates onto the subspace \(Aa=0\). We use \(d\) from \eqref{eq:theory-economic-gap}.

For \(j\in\{S,U,E\}\), let \(\mathcal R_j=\mathbb E[\mathcal R(f_{\hat\beta_j})\mid\mathcal G]\), where \(\hat\beta_S=0\) is the frozen fit's update. Write \(v_U=\operatorname{tr}(MV_U)\) and \(v_E=\operatorname{tr}(MPV_UP^\top)\) for the ordinary and structural fits' prediction variances under \(Q\).

\begin{theorem}[Expected prediction gains from economic equalities]\label{thm:theory-expected-gains}
The conditional expected gains under \mbox{Assumptions~\ref{ass:theory-regularity} and \ref{ass:theory-equalities}} are
\begin{align}
 \mathcal R_S-\mathcal R_E
  &=\|\beta_*\|_M^2-\|Pe+Jd\|_M^2-v_E,\label{eq:theory-gain-freezing}\\
 \mathcal R_U-\mathcal R_E
  &=v_U-v_E+\|e\|_M^2-\|Pe+Jd\|_M^2.\label{eq:theory-gain-ordinary}
\end{align}
In particular, the structural fit strictly improves on both fits exactly when
\[
 \|Pe+Jd\|_M^2<
 \min\bigl\{\|\beta_*\|_M^2-v_E,\ \|e\|_M^2+v_U-v_E\bigr\}.
\]
\end{theorem}

In \eqref{eq:theory-gain-freezing}, \(\|\beta_*\|_M^2\) is the reduction in prediction error from the best affine update. Structural fine-tuning beats freezing if its squared bias and variance are smaller than this reduction. For the comparison with ordinary fine-tuning in \eqref{eq:theory-gain-ordinary}, the variance difference \(v_U-v_E\) can be positive or negative. If \(V_U=cH^{-1}\) for \(c\geq0\), imposing the equalities cannot increase prediction variance. This relation holds, for example, with homoskedastic uncorrelated errors, \(K=G\), and positive definite \(G\). We then have \(V_U-PV_UP^\top=cH^{-1}A^\top(AH^{-1}A^\top)^{-1}AH^{-1}\) positive semidefinite, so \(v_E\le v_U\) for every \(M\). 

When \(e=0\), the ordinary fit is conditionally unbiased for the best update, and imposing the equalities adds \(\|Jd\|_M^2\) of squared bias. Structural fine-tuning then beats ordinary fine-tuning if \(\|Jd\|_M^2<v_U-v_E\), and it beats freezing if \(\|Jd\|_M^2+v_E<\|\beta_*\|_M^2\).

Importantly, we can use Theorem~\ref{thm:theory-expected-gains} to choose economic restrictions. For a proposed equality \(T(f)=\ell\), the gain over ordinary fine-tuning is the variance saved under \(Q\) minus the increase in squared bias. Economic theory or independent data may give a bound on the error in \(\ell\). This bound lets us test the preceding sufficient condition in Theorem \ref{thm:theory-expected-gains}.\footnote{As a special case of Theorem~\ref{thm:theory-expected-gains}, suppose \(G=M=K=I_p\), \(h_X=0\), \(b=0\), and \(\Sigma=\sigma^2 I_n\). The structural fit beats freezing exactly when \(\|P\beta_*\|_2^2>(p-q)\sigma^2/[n(1+2\alpha)]\), and beats ordinary fine-tuning exactly when \(\|(I_p-P)\beta_*\|_2^2<q\sigma^2/[n(1+2\alpha)]\).}

\subsection{Convex economic restrictions}\label{sec:theory-convex-restrictions}

We usually express the economic restrictions as convex inequalities. In the household example, we require predicted shares to be nonincreasing and predicted spending to be concave in budget. On a fixed budget grid, these are linear inequalities in \(\beta\), since \(f_\beta\) is affine in \(\beta\).

For nonnegative last-layer gates, the ordinary and structural coefficient sets are
\[
 B=\{\beta\in\mathbb R^p:\beta\geq-\mathbf1_p\},
 \qquad C=\{\beta\in B:f_\beta\in\mathcal C_E\}.
\]
For this comparison, let \(\hat\beta_U\) and \(\hat\beta_E\) minimize \eqref{eq:theory-coefficient-criterion} over \(B\) and \(C\), respectively, using the same target sample and \(\alpha\). Write \(\hat f_j=f_{\hat\beta_j}\), \(j\in\{U,E\}\).

\begin{corollary}[Convex economic restrictions]\label{cor:theory-convex-restrictions}
Suppose \(\hat f_S,m_T,\phi_1,\ldots,\phi_p\in L^2(Q)\), \(M\) is positive definite, \(H=cM\) for some \(c>0\), and \(C\) is closed and convex with \(\beta_*\in C\). Let \(\tilde\beta=H^{-1}D^\top Z/n\) be the minimizer without gate or economic constraints, and let \(r=\tilde\beta-\hat\beta_U\). For every realization of the target outcomes,
\begin{equation}\label{eq:theory-convex-gain}
 \mathcal R(\hat f_U)-\mathcal R(\hat f_E)
 \geq \|\hat f_U-\hat f_E\|_Q^2
       -2r^\top M(\hat\beta_E-\beta_*).
\end{equation}
If \(r^\top M(\hat\beta_E-\beta_*)\leq0\), the gain is at least \(\|\hat f_U-\hat f_E\|_Q^2\). In particular, \(\tilde\beta\in B\) gives \(r=0\).
\end{corollary}

With \(H=cM\), the fitting criterion and prediction error under \(Q\) weight coefficient changes in the same way, up to a positive constant. This holds, for example, if \(G=K=M\). We allow the best affine approximation to differ from \(m_T\), but require \(\beta_*\in C\), so it satisfies the gate bounds and economic restrictions. When the gate bounds change the ordinary fit, the projection of \(\tilde\beta\) onto \(C\) can lie farther from \(\beta_*\) in the \(M\)-norm than the projection onto \(B\). In Appendix~\ref{appendix:proofs}, adding a valid economic restriction raises prediction error from \(17\) to \(18\).

\section{Simulation}\label{sec:simulation}

In this simulation exercise, we study whether structural fine-tuning improves an issuer's credit limit decisions after an economic shift. Motivated by the 2014 Medicaid expansion, we introduce an exogenous medical payment program that pays one-time part of existing qualifying medical bills for eligible borrowers. Holding borrower composition fixed, this policy changes their conditional probability of delinquency. The lender has a historical risk predictor trained from pre-treatment period and observes outcomes for a small sample of borrowers under the new economy. It then adapts its predictor to grant credit limit increases in order to maximize its profits. A mismatch can generate credit losses due to account delinquency, and a missed profitable increase forgoes interest and fee income.

The experiment compares the frozen source predictor, ordinary fine-tuning, and fine-tuning with economic restrictions. We report the performance of each restriction separately and altogether. Our main outcome is annual credit issuer profit. We also evaluate prediction accuracy and time efficiency.

\subsection{Data Generating Process}\label{sec:sim-dgp}

We simulate 32,768 borrowers histories \(X\) with 1,000 covariates, including credit characteristics, medical payment information, and noisy proxy features. Let \(m_S(X)\) and \(m_T(X)\) denote the source and target probabilities of severe delinquency. The target environment differs from the source, as the medical payment program changes delinquency risk by reducing eligible borrowers' medical liabilities.

Let \(E_i\) and \(B_i\) indicate eligibility and an outstanding medical bill, respectively, each drawn independently from \(\operatorname{Bernoulli}(0.5)\). Then, medical liability and the program payment are
\[
M_i=B_i(0.5+2.5v_i),\qquad
P_i=E_iM_i(0.25+0.75c_i),
\]
where \(v_i,c_i\) are i.i.d. uniform draws on \([0,1]\). Let \(G_i(m)\) denote the maximum liquidity shortfall across these periods when outstanding medical liability is \(m\). The target delinquency risk satisfies
\begin{equation}\label{eq:sim-target-risk}
\operatorname{logit}m_T(X_i)
=\operatorname{logit}m_S(X_i)
-4\{G_i(M_i)-G_i(M_i-P_i)\},
\qquad
Y_i\mid X_i\sim\operatorname{Bernoulli}(m_T(X_i)).
\end{equation}
Thus, medical payments weakly reduce delinquency risk while borrowers receiving no payment are unaffected.

Finally, the credit issuer uses the adapted model to choose credit-limit increases of \(a\in\{0,0.25\}\). Annual issuer profit consists of interest and interchange revenue minus funding costs, credit losses, and servicing expenses. We evaluate how structural fine-tuning affects both risk prediction and the profits generated.

\subsection{Economic Restrictions}\label{sec:sim-restrictions}

We consider seven economic restrictions on how medical payments affect delinquency risk.

First, the \textit{eligibility restriction} requires that risk remain unchanged for ineligible borrowers. Second, the \textit{no-bill restriction} imposes the same condition for borrowers without qualifying medical liabilities:
\begin{align}
m_T(X_i)&=m_S(X_i) &&\text{if }E_i=0,\label{eq:sim-eligibility}\\
m_T(X_i)&=m_S(X_i) &&\text{if }B_i=0.\label{eq:sim-no-bill}
\end{align}

Third, the \textit{cash-flow restriction} requires that, among borrowers receiving payments, target risk depend on borrower history only through historical risk and relevant cash-flow characteristics. Let \(Z_i\) summarize utilization, medical liabilities, coverage, and cash-flow gaps, and define
\[
F(s,z)=\mathbb E_{P_T}[Y_i\mid m_S(X_i)=s,Z_i=z,E_i=B_i=1].
\]
We impose
\begin{equation}\label{eq:sim-cash-restriction}
|m_T(X_i)-F(m_S(X_i),Z_i)|
\leq\epsilon,\qquad E_i=B_i=1,
\end{equation}
where tolerance \(\epsilon=0.02\) allows limited deviations.

The remaining four restrictions bound group-level delinquency risk. For a predetermined borrower group \(\mathcal G_k\), let
\[
\mu_{0k}=\mathbb E_T[m_S(X_i)\mid \mathcal G_k],
\qquad
\mu_{1k}=\mathbb E_T[m_T(X_i)\mid \mathcal G_k],
\]
and let \(q_k\) denote the affected share
$$q_k = \Pr_T(P_i>0|X_i\in \mathcal{G}_k).$$
Define the nonnegative risk-index reduction
\[
\Delta_i=4\{G_i(M_i)-G_i(M_i-P_i)\},
\qquad 0\leq\Delta_i\leq\bar d,
\]
and \(B_k^{\mathrm{sav}}=\mathbb E_T[\tanh(\Delta_i/4)\mid\mathcal G_k]\).

The four group-level restrictions are
\begin{align}
\mu_{1k}
&\leq \mu_{0k},
&&\text{(No-harm bound)}
\label{eq:sim-no-harm}\\
\mu_{1k}
&\geq \mu_{0k}-q_k\tanh(\bar d/4),
&&\text{(Exposure-and-savings cap)}
\label{eq:sim-exposure}\\
\mu_{1k}
&\geq \sigma\!\left(\operatorname{logit}(\mu_{0k})-\bar d\right),
&&\text{(Residual-risk floor)}
\label{eq:sim-risk-floor}\\
\mu_{1k}
&\geq \mu_{0k}-B_k^{\mathrm{sav}},
&&\text{(Savings-distribution bound)}
\label{eq:sim-savings-bound}
\end{align}
where \(\sigma\) denotes the logistic function.

\subsection{Fine-tuning and the issuer's decision rule} \label{sec:sim-methods}

We apply three fine-tuning methods to the pretrained neural network: neuron gating, rank-one LoRA, and full tuning. These methods update 24,576, 40,960, and approximately 88 million parameters, respectively. We use \eqref{eq:method-estimator} with penalty $\alpha = 1$ as the fine-tuning objective. Each trial of simulation includes 1,000 target borrowers with 800 for fitting and 200 for validation. 

For credit limit decisions, let \(b, r\) denote bureau characteristics and predicted delinquency risk. We estimate profit functions \(\hat g_a(b,r)\) using a standalone sample generated before the policy with randomized credit limit increases. Then, after the medical payment policy is implemented, for trial $s$ and fine-tuning method $j$, the credit issuer chooses between a 25\% credit limit increase and no change:
\begin{equation}\label{eq:sim-offer-rule}
\begin{aligned}
a_{f_{js}}^*(X_{is})
&\in \arg\max_{a\in\{0,0.25\}}
      \hat g_a\bigl(b_{is},f_{js}(X_{is})\bigr).
\end{aligned}
\end{equation}
Note that the issuer does not directly observe medical payment records or eligibility. This policy only affects issuer's decision through altered delinquency risk.

\subsection{Evaluation}\label{sec:sim-evaluation}

We use ten Monte Carlo simulation with independent target samples and evaluation samples of \(N=32{,}768\) borrowers. We employ the same pretrained risk predictor and profit model across simulations. The issuer determines credit limit increases before observing evaluation outcomes. For fine-tuning method \(j\), average annual profit per borrower in simulation \(s\) is
\[
V_{js}=\frac1N\sum_{i=1}^N\Pi_{is}\{a_{f_{js}}^*(X_{is})\}.
\]

Our primary outcome is policy regret, defined as expected annual profit forgone relative to an oracle allocation. Let \(\bar\Pi_{is}(a)\) denote borrower-specific expected profit under decision \(a\), averaging over future shocks. Regret is
\[
R_{js}
=
\frac1N\sum_{i=1}^N
\left[
\max_{a\in\{0,0.25\}}\bar\Pi_{is}(a)
-
\bar\Pi_{is}\{a_{f_{js}}^*(X_{is})\}
\right].
\]
We report regret reductions in dollars values r and percentages, averaged across all borrowers, and we compare structural fine-tuning with ordinary fine-tuning and the frozen source predictor. We also evaluate prediction accuracy by RMSE against the true target delinquency probabilities.

\subsection{Results}\label{sec:sim-results}
Table~\ref{tab:sim-main-results} shows that structural fine-tuning with all seven restrictions reduce regret by 6.46\%, 10.21\%, and 3.13\% relative to ordinary neuron gating, LoRA, and full tuning, respectively. RMSE falls by 28.15\%, 29.35\%, and 38.72\%. Gating, LoRA, and full tuning gives similar RMSEs in Panel C. The median time to fit is 14.8, 14.5, and 39.5 seconds. 

\begin{table}[htbp]
\centering
\begin{threeparttable}
\caption{Forecast performance, policy regret, and fitting cost}
\label{tab:sim-main-results}
\begingroup
\footnotesize
\setlength{\tabcolsep}{3pt}
\renewcommand{\arraystretch}{1.12}
\begin{tabular*}{\textwidth}{@{\extracolsep{\fill}}lrrrrrrrr@{}}
\toprule
& \multicolumn{3}{c}{Policy regret (\$)}
& \multicolumn{3}{c}{Risk RMSE (pp)}
& \shortstack{Median fit\\(seconds)} & \shortstack{Parameters\\updated} \\
\cmidrule(lr){2-4}\cmidrule(lr){5-7}
Estimator & Mean & \shortstack{vs.\ frozen\\(\%)}
& \shortstack{vs.\ ordinary\\(\%)}
& Mean & \shortstack{vs.\ frozen\\(\%)}
& \shortstack{vs.\ ordinary\\(\%)} & & \\
\midrule
\multicolumn{9}{@{}l}{\textit{A. Frozen historical model}}\\
Calibrated frozen & 3.550 & --- & --- & 8.180 & --- & --- & --- & 0\\
\addlinespace[4pt]
\multicolumn{9}{@{}l}{\textit{B. Ordinary fine-tuning}}\\
Neuron gating & 3.310 & 6.78 & --- & 7.646 & 6.52 & --- & 22.7 & 24,576\\
Rank-one LoRA & 3.440 & 3.12 & --- & 7.823 & 4.36 & --- & 22.1 & 40,960\\
Full tuning & 3.211 & 9.56 & --- & 8.937 & $-9.26$ & --- & 37.0 & $88$ million\\
\addlinespace[4pt]
\multicolumn{9}{@{}l}{\textit{C. All seven economic restrictions}}\\
Neuron gating & 3.096 & 12.80 & 6.46 & 5.494 & 32.84 & 28.15 & 14.8 & 24,576\\
Rank-one LoRA & 3.088 & 13.01 & 10.21 & 5.527 & 32.43 & 29.35 & 14.5 & 40,960\\
Full tuning & 3.110 & 12.39 & 3.13 & 5.476 & 33.05 & 38.72 & 39.5 & $88$ million\\
\bottomrule
\end{tabular*}
\endgroup
\begin{tablenotes}[flushleft]
\footnotesize
\item \textit{Notes:} Policy regret and RMSE are compared against frozen model and ordinary fine-tuning. Panel C uses $\epsilon=0.02$, $\delta=0$. Parameter updated refer to the number of parameters fine-tuned for each methods.
\end{tablenotes}
\end{threeparttable}
\end{table}

In Table~\ref{tab:sim-tolerance}, we vary how far estimated risk can move from the fitted cash-flow center for paid borrowers ($\epsilon$) or from calibrated historical risk for unpaid borrowers ($\delta$). Comparisons use $\epsilon=\delta=0$. Positive values of $\epsilon$ modestly lower regret relative to $\epsilon=0$, with little change in RMSE. With $\epsilon=0$, setting $\delta=0.05$ reduces regret by 4.53\%, 3.30\%, and 5.36\%, while increasing RMSE by 5.34\%, 5.12\%, and 5.03\%. Thus, tighter predictive restrictions need not improve credit allocation. Also, we separately estimate the contribution of each ER in Appendix.

\begin{table}[htbp]
\centering
\begin{threeparttable}
\caption{Sensitivity to economic tolerances}
\label{tab:sim-tolerance}
\begingroup
\small
\setlength{\tabcolsep}{4pt}
\renewcommand{\arraystretch}{1.14}
\begin{tabular*}{\textwidth}{@{\extracolsep{\fill}}lrrrrrr@{}}
\toprule
& \multicolumn{2}{c}{Neuron gating}
& \multicolumn{2}{c}{Rank-one LoRA}
& \multicolumn{2}{c}{Full tuning}\\
\cmidrule(lr){2-3}\cmidrule(lr){4-5}\cmidrule(lr){6-7}
Tolerance & Mean & \shortstack{Reduction (\%)\\(paired SE)}
& Mean & \shortstack{Reduction (\%)\\(paired SE)}
& Mean & \shortstack{Reduction (\%)\\(paired SE)}\\
\midrule
\multicolumn{7}{@{}l}{\textit{A. Policy regret (\$): vary $\epsilon$, with $\delta=0$}}\\
0 & 3.154 & --- & 3.156 & --- & 3.147 & ---\\
0.02 & 3.096 & $+1.84\;(1.03)$ & 3.088 & $+2.15\;(1.02)$ & 3.110 & $+1.17\;(0.62)$\\
0.05 & 3.085 & $+2.19\;(1.20)$ & 3.109 & $+1.50\;(0.98)$ & 3.112 & $+1.12\;(1.02)$\\
0.10 & 3.102 & $+1.64\;(0.93)$ & 3.138 & $+0.57\;(0.74)$ & 3.098 & $+1.57\;(0.82)$\\
\addlinespace[6pt]
\multicolumn{7}{@{}l}{\textit{B. Risk RMSE (pp): vary $\epsilon$, with $\delta=0$}}\\
0 & 5.493 & --- & 5.508 & --- & 5.466 & ---\\
0.02 & 5.494 & $-0.02\;(0.18)$ & 5.527 & $-0.36\;(0.37)$ & 5.476 & $-0.19\;(0.30)$\\
0.05 & 5.501 & $-0.16\;(0.43)$ & 5.578 & $-1.28\;(0.55)$ & 5.504 & $-0.69\;(0.58)$\\
0.10 & 5.492 & $+0.01\;(0.79)$ & 5.588 & $-1.45\;(0.98)$ & 5.593 & $-2.32\;(1.25)$\\
\addlinespace[6pt]
\multicolumn{7}{@{}l}{\textit{C. Policy regret (\$): vary $\delta$, with $\epsilon=0$}}\\
0 & 3.154 & --- & 3.156 & --- & 3.147 & ---\\
0.02 & 3.048 & $+3.37\;(1.40)$ & 3.123 & $+1.06\;(1.80)$ & 3.106 & $+1.32\;(2.12)$\\
0.05 & 3.011 & $+4.53\;(2.38)$ & 3.052 & $+3.30\;(2.22)$ & 2.979 & $+5.36\;(2.35)$\\
0.10 & 3.062 & $+2.91\;(2.16)$ & 3.096 & $+1.92\;(2.19)$ & 3.054 & $+2.96\;(1.56)$\\
\addlinespace[6pt]
\multicolumn{7}{@{}l}{\textit{D. Risk RMSE (pp): vary $\delta$, with $\epsilon=0$}}\\
0 & 5.493 & --- & 5.508 & --- & 5.466 & ---\\
0.02 & 5.582 & $-1.63\;(0.92)$ & 5.604 & $-1.75\;(0.73)$ & 5.550 & $-1.54\;(0.69)$\\
0.05 & 5.786 & $-5.34\;(2.39)$ & 5.789 & $-5.12\;(2.09)$ & 5.741 & $-5.03\;(2.63)$\\
0.10 & 5.783 & $-5.29\;(3.48)$ & 5.888 & $-6.91\;(3.04)$ & 5.766 & $-5.48\;(2.77)$\\
\bottomrule
\end{tabular*}
\endgroup
\begin{tablenotes}[flushleft]
\footnotesize
\item \textit{Notes:} Reductions are relative to $\epsilon=\delta=0$ within each family; parentheses contain paired Monte Carlo standard errors. Positive reductions indicate improvement.
\end{tablenotes}
\end{threeparttable}
\end{table}

\clearpage
\section{Conclusion}\label{sec:conclusion}

This paper develops a framework of structural fine-tuning that incorporates economic restrictions into the adaptation of pretrained predictors. Without requiring a fully specified structural model, our approach uses economic knowledge to guide fine-tuning with limited target data. We characterize the conditions under which these restrictions improve prediction relative to freezing the source model or ordinary fine-tuning. In particular, we investigate the trade-off between estimation variance and restriction-induced bias.

In a simulated Medicaid-style medical payment program, structural fine-tuning reduces prediction error and policy regret across neuron gating, LoRA, and full fine-tuning. The improvements are primarily driven by cash-flow restrictions, while group-level bounds provide limited additional gains. These results illustrate the potential of economic knowledge to improve model adaptation and downstream decisions. More broadly, the framework provides a flexible approach to incorporating economic theory into pretrained models without sacrificing their predictive flexibility.

\clearpage
\appendix
\input{appendix}

\clearpage
\bibliographystyle{chicago}
\bibliography{reference}

\end{document}

%% file: appendix.tex
\section*{Appendix}
\section{Proofs}\label{appendix:proofs}\label{appendix}
\begin{proof}[Proof of Theorem~\ref{thm:theory-expected-gains}]
All expectations and variances in this proof are conditional on \(\mathcal G\). Let \(z=D^\top Z/n\) and \(B_A=AH^{-1}A^\top\). Assumption~\ref{ass:theory-regularity} gives \(B_A\succ0\).

Up to a term independent of \(\beta\), the criterion in \eqref{eq:theory-coefficient-criterion} is \(\beta^\top H\beta-2z^\top\beta\). Since \(H\succ0\), the ordinary fit solves \(H\hat\beta_U=z\). The constrained first-order conditions, with multiplier \(\lambda\in\mathbb R^q\), are
\[
 H(\hat\beta_E-\hat\beta_U)=A^\top\lambda,\qquad A\hat\beta_E=b.
\]
Substituting \(\hat\beta_E=\hat\beta_U+H^{-1}A^\top\lambda\) into the equality gives
\begin{align*}
 B_A\lambda&=b-A\hat\beta_U,\\
 \hat\beta_E&=\hat\beta_U+H^{-1}A^\top B_A^{-1}(b-A\hat\beta_U)
             =P\hat\beta_U+Jb.
\end{align*}
Strict convexity makes this solution the unique constrained minimizer.

Let \(\eta=H^{-1}D^\top\xi/n\). Equation~\eqref{eq:theory-design-noise} gives \(\mathbb E[\eta]=0\) and \(\mathbb E[\eta\eta^\top]=V_U\). Using \(Z=D\beta_*+h_X+\xi\), we have
\begin{align*}
 \hat\beta_U-\beta_*
 &=H^{-1}\{(G-H)\beta_*+D^\top h_X/n\}+\eta
   =e+\eta,\\
 \hat\beta_E-\beta_*
 &=P(\hat\beta_U-\beta_*)+J(b-A\beta_*)
   =Pe+Jd+P\eta.
\end{align*}
For any \(\mathcal G\)-measurable vector \(a\) and matrix \(C\) of compatible dimensions,
\begin{align*}
 \mathbb E\|a+C\eta\|_M^2
 &=\|a\|_M^2+2a^\top MC\mathbb E[\eta]
   +\operatorname{tr}\{MC\mathbb E[\eta\eta^\top]C^\top\}\\
 &=\|a\|_M^2+\operatorname{tr}(MCV_UC^\top).
\end{align*}
Applying this identity to \eqref{eq:theory-affine-risk} yields
\begin{align*}
 \mathcal R_S &= \|h\|_Q^2+\|\beta_*\|_M^2,\\
 \mathcal R_U &= \|h\|_Q^2+\|e\|_M^2+v_U,\\
 \mathcal R_E &= \|h\|_Q^2+\|Pe+Jd\|_M^2+v_E.
\end{align*}
Subtracting \(\mathcal R_E\) gives \eqref{eq:theory-gain-freezing} and \eqref{eq:theory-gain-ordinary}. Both gains are positive exactly when \(\|Pe+Jd\|_M^2\) is less than the minimum in the theorem.

When \(V_U=cH^{-1}\) with \(c\geq0\), the identities \(AH^{-1}=B_A J^\top\) and \(P=I_p-JA\) give
\begin{align*}
 PH^{-1}P^\top
 &=H^{-1}-JAH^{-1}-H^{-1}A^\top J^\top+JB_A J^\top\\
 &=H^{-1}-JB_A J^\top,\\
 v_U-v_E
 &=c\,\operatorname{tr}\!\left(B_A^{1/2}J^\top MJ B_A^{1/2}\right)
 \geq 0.
\end{align*}
The last inequality follows from \(M\succeq0\). If \(K=G\) and \(\Sigma=\sigma^2 I_n\), then \(H=(1+\alpha)G\), so
\[
 V_U=\frac{\sigma^2}{n}H^{-1}GH^{-1}
   =\frac{\sigma^2}{n(1+\alpha)}H^{-1}.
\]

For \(j\in\{S,U\}\), let \(g_j=\mathcal R_j-\mathcal R_E\). The gain \(g_j^0\) uses \(b=A\beta_*\) with \(A,H,e,V_U,M\) held fixed. If \(\|Jd\|_M\leq\varepsilon\), then
\begin{align*}
 g_j^0-g_j
 &=\|Pe+Jd\|_M^2-\|Pe\|_M^2\\
 &=2(Pe)^\top MJd+\|Jd\|_M^2\\
 &\leq 2\|Pe\|_M\,\varepsilon+\varepsilon^2,
\end{align*}
by Cauchy--Schwarz applied after multiplication by \(M^{1/2}\). Hence both gains are positive if this bound is less than \(\min\{g_S^0,g_U^0\}\).

\begin{samepage}
In the orthonormal case \(G=M=K=I_p\), \(h_X=0\), \(b=0\), and \(\Sigma=\sigma^2 I_n\), set \(t=(1+\alpha)^{-1}\). Then \(P=I_p-A^\top(AA^\top)^{-1}A\) is the Euclidean projection onto \(\ker A\), which has dimension \(p-q\), and
\begin{align*}
 e&=-(1-t)\beta_*,&
 Jd&=-(I_p-P)\beta_*,&
 V_U&=t^2\sigma^2 I_p/n,\\
 v_U&=t^2p\sigma^2/n,&
 v_E&=t^2(p-q)\sigma^2/n.
\end{align*}
\end{samepage}
Since \(P\beta_*\) and \((I_p-P)\beta_*\) are orthogonal,
\[
 \|Pe+Jd\|_2^2=(1-t)^2\|P\beta_*\|_2^2+\|(I_p-P)\beta_*\|_2^2.
\]
Substitution into \eqref{eq:theory-gain-freezing} and \eqref{eq:theory-gain-ordinary} yields
\begin{align*}
 \mathcal R_S-\mathcal R_E
 &=t(2-t)\|P\beta_*\|_2^2-t^2(p-q)\sigma^2/n,\\
 \mathcal R_U-\mathcal R_E
 &=t^2q\sigma^2/n-t(2-t)\|(I_p-P)\beta_*\|_2^2.
\end{align*}
Dividing by \(t(2-t)>0\) and using \(t/(2-t)=1/(1+2\alpha)\) gives the two thresholds.
\end{proof}

For an example in which imposing the restriction increases \(Q\)-weighted variance, take \(n=1\), \(D=(0,2)\), \(K=M=I_2\), \(\alpha=1\), and \(\Sigma=(\sigma^2)\) with \(\sigma^2>0\). Impose \(\beta_1=\sqrt{3}\beta_2\), and take \(\beta_*=0\) and \(h=0\), so the restriction is correct. The ordinary fit is \((0,2\xi/5)^\top\), giving \(v_U=4\sigma^2/25\). In the restricted problem, write \(\beta=s(\sqrt{3}/2,1/2)^\top\). The criterion is \((\xi-s)^2+s^2\), so \(v_E=\sigma^2/4>v_U\).

\begin{proof}[Proof of Corollary~\ref{cor:theory-convex-restrictions}]
Since \(H=cM\), completing the square in \eqref{eq:theory-coefficient-criterion} gives
\[
 \frac1n\|Z-D\beta\|_2^2+\alpha\beta^\top K\beta
 =c\|\beta-\tilde\beta\|_M^2
   +\frac{Z^\top Z}{n}-\tilde\beta^\top H\tilde\beta.
\]
The sets \(B\) and \(C\) are nonempty, closed, and convex. Since \(M\) is positive definite, the two minimizers are unique and satisfy
\[
 \hat\beta_U=\underset{\beta\in B}{\arg\min}
                 \|\beta-\tilde\beta\|_M^2,
 \qquad
 \hat\beta_E=\underset{\beta\in C}{\arg\min}
                 \|\beta-\tilde\beta\|_M^2.
\]
For \(t\in[0,1]\), convexity and \(\beta_*\in C\) give
\(\hat\beta_E+t(\beta_*-\hat\beta_E)\in C\). Hence
\begin{align*}
 0
 &\leq \left.\frac{d}{dt}
       \|\hat\beta_E+t(\beta_*-\hat\beta_E)-\tilde\beta\|_M^2
       \right|_{t=0+}\\
 &=2(\hat\beta_E-\tilde\beta)^\top M(\beta_*-\hat\beta_E).
\end{align*}
Using \(\tilde\beta=\hat\beta_U+r\), this is
\[
 (\hat\beta_U-\hat\beta_E)^\top M(\hat\beta_E-\beta_*)
 \geq-r^\top M(\hat\beta_E-\beta_*).
\]
By \eqref{eq:theory-affine-risk},
\begin{align*}
 \mathcal R(\hat f_U)-\mathcal R(\hat f_E)
 &=\|\hat\beta_U-\beta_*\|_M^2-\|\hat\beta_E-\beta_*\|_M^2\\
 &=\|\hat\beta_U-\hat\beta_E\|_M^2
   +2(\hat\beta_U-\hat\beta_E)^\top M(\hat\beta_E-\beta_*)\\
 &\geq\|\hat f_U-\hat f_E\|_Q^2
       -2r^\top M(\hat\beta_E-\beta_*).
\end{align*}
If \(r^\top M(\hat\beta_E-\beta_*)\leq0\), the last term is nonnegative. If \(\tilde\beta\in B\), the unique minimizer over \(B\) is \(\tilde\beta\), so \(r=0\).
\end{proof}

The correction in \eqref{eq:theory-convex-gain} cannot generally be dropped. Take \(M=G=K=I_2\), \(\alpha=0\), \(\tilde\beta=(-2,2)^\top\), \(\beta_*=(3,3)^\top\), and \(h=0\). Let \(C=B\cap\{\beta:\beta_1\geq\beta_2\}\). Then \(\beta_*\in C\), but \(\hat\beta_U=(-1,2)^\top\) and \(\hat\beta_E=(0,0)^\top\). The prediction errors are \(17\) and \(18\), respectively, so the structural fit has larger error for this target sample.

\section{Details of monte carlo simulation}

\subsection{Data Generating Process}\label{app:v8-credit-dgp}

Let $P_S$ and $P_T$ denote the joint distributions of $(X,Y)$ before and after Medicaid expansion. We generate the covariates in the same way in both regimes. For $e=S,T$, $Y_e$ records severe delinquency within twelve months and $m_e(X)=\Pr(Y_e=1\mid X)$. It is the conditional risk, not the distribution of the covariates, that changes.

The observed borrower record is $W=(s,u,n,\ell,d)$. The tier $s$ takes the values 0, 1, and 2, with shares 0.30, 0.45, and 0.25. Utilization $u$ and number of accounts $n$ are drawn independently at a given tier. For the former the support is $\mathcal U=(0,0.50,0.95,1.10)$ and for the latter $\mathcal N=(1,3,6)$. The relative probabilities, in support order, are

$$
 \begin{aligned}
 \Pr(u=v\mid s)&\propto\omega_v e^{-0.8sv}, & \omega&=(0.30,0.35,0.25,0.10),\\
 \Pr(n=k\mid s)&\propto\xi_k k^{0.35(s-1)}, & \xi&=(0.40,0.40,0.20).
 \end{aligned}
$$

We draw the age of the credit file $\ell$ independently of these variables. It is 24, 96, or 240 months with probabilities $(0.25,0.50,0.25)$. We draw $d$, the indicator for prior severe delinquency, with probability $\sigma(-1.92355+1.4u-0.6s)$ of one conditional on the other four fields. Here $\sigma$ is the logistic function and $\operatorname{logit}(p)=\log\{p/(1-p)\}$. The normalized credit limit is $l_0=0.10+0.12s+0.03n$, so the normalized balance is $b_0=ul_0$. With $[x]_+=\max\{x,0\}$, the credit report is

$$
 b=\big(s,u,n,\ell,d,560+90s-30d-40u,l_0,b_0,[l_0-b_0]_+,
            b_0+0.08n,d(1+\mathbf 1\{n=6\}),\mathbf 1\{u>1\}\big).
$$

We draw 32 covariates $R_k$ and 956 innovations $e_j$ independently from $U(-\sqrt3,\sqrt3)$. Write $R=(R_1,\ldots,R_{32})$ and set $v_k=(R_k/\sqrt3+1)/2$ and $z=(s-1,u,n/3,\ell/120,d)$. We use $X=(b,R,X^{\mathrm{proxy}})\in\mathbb R^{1000}$, with proxy variables

$$
 X^{\mathrm{proxy}}_j=\frac{\sum_{k=1}^5 z_k\sin(kj)}{\sqrt5}
   +\frac{0.6\sum_{k=1}^{32}R_k\cos(kj)}{\sqrt{32}}+0.2e_j,\qquad j=1,\ldots,956.
$$

We assign each borrower an unobserved financial type $k\in\{0,1,2\}$. Given $W$, the type probabilities are proportional to the exponentials of $(0,-0.2+0.5u+0.5d-0.2s,-0.8+0.8u+0.8d-0.4s)$. Set $h=0.4$ and $F=\exp(-0.1+0.45s-0.4k)$. For $\rho$, the fraction of medical spending paid out of pocket, we distinguish two cases. At $F\leq0.5$, take $\rho=0.10$. Otherwise put probability $\sigma(-1+0.8s-0.4k)$ on $\rho=0.30$ and the rest on $\rho=0.65$. The same $W,k$ gives a fresh draw for the budget state $j$, independent of $\rho$. Its probabilities for $0,1,2$ are $0.85-0.10k$, $0.12+0.06k$, and $0.03+0.04k$. These states amount to the following choices for income, medical spending, and other obligations $(I_j,M_j^S,O_j)$.

$$
 h(F,0,0.02),\qquad h(0.9F,0.3,0.04),\qquad h(0.6F,1.5,0.10).
$$

For a borrower in state $\nu=(k,\rho,j)$ the card payment is $p_c=0.25b_0+0.015n$ and cash assets are $A=0.10h\exp(0.3s-0.5k)$. Then $p_\nu$ is the risk at that state. The net obligations enter the log odds as follows.

$$
\begin{aligned}
 \operatorname{logit}p_\nu
  &=-3.56253+3.32639d+0.35k+0.30u+0.10\log(1+n)\\
  &\quad +\frac{p_c+\rho M^S_j+O_j-A-I_j+0.60h}{0.25h}.
\end{aligned}
$$

Write $\bar p_W=\mathbb E[p_\nu\mid W]$ and $v_W=\mathbb E[p_\nu(1-p_\nu)\mid W]$. We set the historical risk to

$$
 m_S(X)=\bar p_W+\tfrac12 v_W\tanh\!\left(\tfrac14\sum_{k=1}^{16}R_{2k-1}R_{2k}\right).
$$

Let $E=\mathbf1\{R_3\geq0\}$ and $B=\mathbf1\{R_4\geq0\}$. The first gives eligibility and the second gives a qualifying bill. We set the bill to $M=B(0.5+2.5v_1)$ and pay $P=EM(0.25+0.75v_2)$ to the creditor. Thus nothing is paid unless both indicators are one.

A borrower need not pay the whole bill in the same cycle. Most of it, 70\%, comes due in cycle $1+\min\{\lfloor6v_{17}\rfloor,5\}$, and 6\% in each of the other five cycles. Let $w_t$ denote these shares. For any liability $m$, the shortfall in the worst cycle is $G(m)$. Compare the worst cycle at the full bill with the worst cycle after payment; the two cycles need not be the same.

$$
 \begin{gathered}
 G(m)=\left[\max_{1\leq t\leq6}\{1.5v_{4+t}+0.35u-1.5v_{10+t}+w_t m\}\right]_+,\\
 \Delta G=G(M)-G(M-P).
 \end{gathered}
$$

An expansion payment reduces delinquency risk only where it reduces the shortfall. The amount $\Delta G$ is determined by $X$. We map it into risk by subtracting $4\Delta G$ from the old log odds.

$$
 m_T(X)=\sigma\!\left(\operatorname{logit}m_S(X)-4\Delta G\right).
$$

We draw $Y_e\mid X\sim\mathrm{Bernoulli}(m_e(X))$ independently across borrowers.

The issuer's profit depends on the amount a borrower spends and repays after an offer. The starting limit and balance are $L=10{,}000l_0$ and $C=10{,}000b_0$. We use $c=\sum_{t=1}^6v_{10+t}/6$ for cash on hand. We draw $U_D,U_R,U_C$ independently from $U(0,1)$ and independently of $X$, and keep the same draws at different offers. The borrower makes one purchase. Under a proportional increase $a$, available credit is $H_a$ and the purchase is the lesser of demand $D$ and available credit. Denote the purchase by $Q_a$, and the differences from $a=0$ in purchases and unused credit by $\Delta Q_a$ and $\Delta H_a$.

$$
\begin{gathered}
 D=L(0.05+0.28u+0.28c+0.48U_D),\qquad
 H_a=[(1+a)L-C]_+,\qquad Q_a=\min\{D,H_a\},\\
 \Delta Q_a=Q_a-Q_0,\qquad \Delta H_a=(H_a-Q_a)-(H_0-Q_0).
\end{gathered}
$$

The borrower repays a fraction $r$ over the year. The closing and average balances are $C_a^{\mathrm{end}}$ and $\bar C_a$, and interest applies to the fraction $\varphi$ of the average balance. We keep $m_e(X)$ fixed across offers. At each offer, given $X,U_D,e$, the account charges off with probability $q_a$.

$$
\begin{gathered}
 r=0.22+0.35c+0.25U_R,\qquad
 C_a^\mathrm{end}=(1-r)(C+Q_a),\qquad \bar C_a=(C+Q_a+C_a^\mathrm{end})/2,\\
 \varphi=\min\{1,[0.15+0.60u+0.16(1-c)]_+\},\qquad J_a=\mathbf 1\{U_C<q_a\},\\
 \operatorname{logit}q_a=\operatorname{logit}m_e(X)-1.05
                      +\frac{2\Delta Q_a-0.65\Delta H_a}{L}.
\end{gathered}
$$

We price a performing loan at par at year end. That changes if the account charges off: the issuer recovers one fifth of remaining principal and collects only 40\% of the normal interest. Interest applies to $\varphi\bar C_a$ and funding costs to $\bar C_a$. Interchange is based on the purchase $Q_a$. The remaining costs in annual dollar profit $\Pi_a$ are for servicing and increasing the limit.

$$
\begin{aligned}
 \Pi_a={}&0.22\,\varphi\bar C_a(1-0.60J_a)+0.015Q_a-0.045\bar C_a
          -0.80J_a C_a^\mathrm{end}\\
        &{}-18-6\,\mathbf 1\{a>0\}-0.001aL .
\end{aligned}
$$

There is one fitted profit function per simulation draw. We fit it using 32,768 historical borrowers, distinct from the 32,768 used to fit the source risk model. In these profit data, assign each $a\in\{0,0.10,0.25\}$ with probability $1/3$. We observe one profit outcome for each borrower, $\Pi_a$ at the assigned limit. Let $p$ be calibrated risk and $u_j^+=[u-j]_+$. Transform the raw regressors into $z_{\mathrm p}(b,p)$ by centering and scaling on the profit sample.

$$
 (n,\ell,d,l_0,b_0,u,u_{0.5}^+,u_1^+,l_0u_{0.5}^+,l_0u_1^+,p,p^2,pu,pd,pl_0,pb_0).
$$

Write the prediction as

$$
 \hat g_a(b,p)=\hat\gamma_a+z_{\mathrm p}(b,p)^\top(\hat\beta+\hat\delta_a).
$$

Here $\hat\gamma_a$ is the intercept for offer $a$, $\hat\beta$ is the common slope, and $\hat\delta_a$ is that offer's slope deviation. The respective ridge weights are $0$, $32$, and $128$.

Three independent samples have 8,192 observations per stratum. The first calibrates the score, the second gives bounds on historical risk, and the third supplies the payment exposure and savings estimates. After calibration and that the historical bounds are fitted, we hold them fixed for the rest of the experiment. For each stratum draw 256 more profiles without outcomes. These are the profiles used for the group restrictions. The bound on the decline in the log odds is $8\Delta G$. The update with the lowest validation log loss on the common splits in the simulation methods is fitted again using all 1,000 pairs.

We use the pre-payment report for the credit offer. When two offers tie, leave the limit as it is. At each $W$ the report-only benchmark averages predicted profit over 32 independent draws. All methods use the same profit fit and the same borrowers in the evaluation sample.

At a given $X$ and $U_D$, average profit first over charge-off and repayment. This amounts to evaluating the profit formula at $J_a=q_a$ and $r=0.22+0.35c+0.25/2$. The expected target profit then integrates over $U_D$. For each comparison between $a$ and 0, cut $[0,1]$ at any distinct interior roots of $D=H_a$ or $D=H_0$. Those are the points where the purchase formula changes. Use 12-point Gauss-Legendre quadrature in each interval. We use seeds 18085001 to 18085010 for training and 18086001 to 18086010 for evaluation.

\subsection{Additional results}

The cash-flow restriction produces the largest standalone improvements in both outcomes (Table~\ref{tab:sim-single-restrictions}). Adding the other restrictions further lowers RMSE for every method, but cash-flow-only gating and full tuning have lower regret than their combined counterparts. Eligibility and no-bill restrictions improve RMSE throughout but increase full-tuning regret. The exposure-and-savings cap and residual-risk floor leave the reported outcomes unchanged.

\begin{table}[htbp]
\centering
\begin{threeparttable}
\caption{Performance when each economic restriction imposed separately}
\label{tab:sim-single-restrictions}
\begingroup
\small
\setlength{\tabcolsep}{3pt}
\renewcommand{\arraystretch}{1.14}
\begin{tabular*}{\textwidth}{@{\extracolsep{\fill}}lrrrrrr@{}}
\toprule
& \multicolumn{2}{c}{Neuron gating}
& \multicolumn{2}{c}{Rank-one LoRA}
& \multicolumn{2}{c}{Full tuning}\\
\cmidrule(lr){2-3}\cmidrule(lr){4-5}\cmidrule(lr){6-7}
Restriction & Mean & \shortstack{Reduction\\(\%)}
& Mean & \shortstack{Reduction\\(\%)}
& Mean & \shortstack{Reduction\\(\%)}\\
\midrule
\multicolumn{7}{@{}l}{\textit{A. Policy regret (dollars per borrower)}}\\
Ordinary & 3.310 & --- & 3.440 & --- & 3.211 & ---\\
Eligibility & 3.260 & \begin{tabular}[c]{@{}r@{}}$+1.50$\\[-2pt]$(1.69)$\end{tabular} & 3.342 & \begin{tabular}[c]{@{}r@{}}$+2.84$\\[-2pt]$(2.34)$\end{tabular} & 3.307 & \begin{tabular}[c]{@{}r@{}}$-3.00$\\[-2pt]$(3.87)$\end{tabular}\\
No-bill & 3.220 & \begin{tabular}[c]{@{}r@{}}$+2.70$\\[-2pt]$(1.61)$\end{tabular} & 3.259 & \begin{tabular}[c]{@{}r@{}}$+5.26$\\[-2pt]$(2.57)$\end{tabular} & 3.341 & \begin{tabular}[c]{@{}r@{}}$-4.05$\\[-2pt]$(3.45)$\end{tabular}\\
Cash-flow & 3.069 & \begin{tabular}[c]{@{}r@{}}$+7.28$\\[-2pt]$(2.07)$\end{tabular} & 3.109 & \begin{tabular}[c]{@{}r@{}}$+9.61$\\[-2pt]$(2.11)$\end{tabular} & 3.054 & \begin{tabular}[c]{@{}r@{}}$+4.89$\\[-2pt]$(3.21)$\end{tabular}\\
No-harm & 3.310 & \begin{tabular}[c]{@{}r@{}}$+0.00$\\[-2pt]$(0.00)$\end{tabular} & 3.392 & \begin{tabular}[c]{@{}r@{}}$+1.37$\\[-2pt]$(1.33)$\end{tabular} & 3.253 & \begin{tabular}[c]{@{}r@{}}$-1.30$\\[-2pt]$(2.25)$\end{tabular}\\
Exposure-and-savings cap & 3.310 & \begin{tabular}[c]{@{}r@{}}$+0.00$\\[-2pt]$(0.00)$\end{tabular} & 3.440 & \begin{tabular}[c]{@{}r@{}}$+0.00$\\[-2pt]$(0.00)$\end{tabular} & 3.211 & \begin{tabular}[c]{@{}r@{}}$+0.00$\\[-2pt]$(0.00)$\end{tabular}\\
Residual-risk floor & 3.310 & \begin{tabular}[c]{@{}r@{}}$+0.00$\\[-2pt]$(0.00)$\end{tabular} & 3.440 & \begin{tabular}[c]{@{}r@{}}$+0.00$\\[-2pt]$(0.00)$\end{tabular} & 3.211 & \begin{tabular}[c]{@{}r@{}}$+0.00$\\[-2pt]$(0.00)$\end{tabular}\\
Savings-distribution bound & 3.310 & \begin{tabular}[c]{@{}r@{}}$+0.00$\\[-2pt]$(0.00)$\end{tabular} & 3.440 & \begin{tabular}[c]{@{}r@{}}$+0.00$\\[-2pt]$(0.00)$\end{tabular} & 3.212 & \begin{tabular}[c]{@{}r@{}}$-0.02$\\[-2pt]$(0.03)$\end{tabular}\\
\addlinespace[6pt]
\multicolumn{7}{@{}l}{\textit{B. Risk RMSE (percentage points)}}\\
Ordinary & 7.646 & --- & 7.823 & --- & 8.937 & ---\\
Eligibility & 7.060 & \begin{tabular}[c]{@{}r@{}}$+7.67$\\[-2pt]$(1.70)$\end{tabular} & 7.206 & \begin{tabular}[c]{@{}r@{}}$+7.90$\\[-2pt]$(2.18)$\end{tabular} & 7.933 & \begin{tabular}[c]{@{}r@{}}$+11.24$\\[-2pt]$(2.79)$\end{tabular}\\
No-bill & 6.942 & \begin{tabular}[c]{@{}r@{}}$+9.22$\\[-2pt]$(2.32)$\end{tabular} & 7.081 & \begin{tabular}[c]{@{}r@{}}$+9.49$\\[-2pt]$(3.45)$\end{tabular} & 7.931 & \begin{tabular}[c]{@{}r@{}}$+11.26$\\[-2pt]$(2.56)$\end{tabular}\\
Cash-flow & 5.955 & \begin{tabular}[c]{@{}r@{}}$+22.12$\\[-2pt]$(5.33)$\end{tabular} & 6.557 & \begin{tabular}[c]{@{}r@{}}$+16.18$\\[-2pt]$(8.90)$\end{tabular} & 6.022 & \begin{tabular}[c]{@{}r@{}}$+32.62$\\[-2pt]$(4.06)$\end{tabular}\\
No-harm & 7.646 & \begin{tabular}[c]{@{}r@{}}$+0.00$\\[-2pt]$(0.00)$\end{tabular} & 7.851 & \begin{tabular}[c]{@{}r@{}}$-0.35$\\[-2pt]$(0.34)$\end{tabular} & 9.414 & \begin{tabular}[c]{@{}r@{}}$-5.33$\\[-2pt]$(3.48)$\end{tabular}\\
Exposure-and-savings cap & 7.646 & \begin{tabular}[c]{@{}r@{}}$+0.00$\\[-2pt]$(0.00)$\end{tabular} & 7.823 & \begin{tabular}[c]{@{}r@{}}$+0.00$\\[-2pt]$(0.00)$\end{tabular} & 8.937 & \begin{tabular}[c]{@{}r@{}}$+0.00$\\[-2pt]$(0.00)$\end{tabular}\\
Residual-risk floor & 7.646 & \begin{tabular}[c]{@{}r@{}}$+0.00$\\[-2pt]$(0.00)$\end{tabular} & 7.823 & \begin{tabular}[c]{@{}r@{}}$+0.00$\\[-2pt]$(0.00)$\end{tabular} & 8.937 & \begin{tabular}[c]{@{}r@{}}$+0.00$\\[-2pt]$(0.00)$\end{tabular}\\
Savings-distribution bound & 7.646 & \begin{tabular}[c]{@{}r@{}}$+0.00$\\[-2pt]$(0.00)$\end{tabular} & 7.823 & \begin{tabular}[c]{@{}r@{}}$+0.00$\\[-2pt]$(0.00)$\end{tabular} & 8.925 & \begin{tabular}[c]{@{}r@{}}$+0.14$\\[-2pt]$(0.14)$\end{tabular}\\
\bottomrule
\end{tabular*}
\endgroup
\begin{tablenotes}[flushleft]
\footnotesize
\item \textit{Notes:} Reductions are relative to ordinary fine-tuning of the same family; parentheses contain paired Monte Carlo standard errors. Positive reductions indicate improvement.
\end{tablenotes}
\end{threeparttable}
\end{table}